\documentclass[a4paper,11pt]{article}
\usepackage{pos}

\title{How to deal with conformal and pure scale-invariant theories of gravity in d dimensions?}

\author*[a]{Anamaria Hell}
\author[b,c]{Dieter L\"ust}

\affiliation[a]{
{Kavli IPMU (WPI), UTIAS, The University of Tokyo,
and Center for Data-Driven Discovery,\\
5-1-5 Kashiwanoha, Kashiwa, Chiba 277-8583, Japan}}

\affiliation[b]{Arnold Sommerfeld Center for Theoretical Physics,
Ludwig–Maximilians–Universität München\\
Theresienstraße 37, 80333 Munich, Germany}

\affiliation[c]{Max–Planck–Institut für Physik (Werner–Heisenberg–Institut)\\
Boltzmannstra{\ss}e 8, 85748 Garching, Germany}

\emailAdd{anamaria.hell@ipmu.jp}
\emailAdd{luest@mpp.mpg.de }

\abstract{ Conformally-invariant and pure, scale-invariant theories of gravity are particularly interesting in four or higher dimensions. Yet, in contrast to their four-dimensional counterparts, theories in higher dimensions are significantly more difficult to study. In these proceedings, following our recent work, we will formulate such theories in d dimensions, present an elegant way to handle them, and show that imposing invariance under scale or conformal transformations gives rise to entirely different properties when compared to their four-dimensional analogues. 
}

\FullConference{Proceedings of the Corfu Summer Institute 2025 "School and Workshops on Elementary Particle Physics and Gravity" (CORFU2025)\\
27 April - 28 September, 2025\\
Corfu, Greece\\}

\begin{document}
\maketitle

\section{Introduction}

One of the most intriguing questions is whether we live in more than four dimensions. This possibility is not only interesting out of sheer curiosity, but is fundamentally important: theoretical models requiring higher dimensions have the potential to resolve some of the long-standing puzzles, such as the unification of gravity and electromagnetism \cite{Kaluza:1921tu, Klein:1926tv}, the hierarchy problem \cite{Arkani-Hamed:1998sfv, Arkani-Hamed:1998jmv, Antoniadis:1998ig, Randall:1999ee, Randall:1999vf}, and provide a (string) theory of quantum gravity \cite{Green:1987sp, Blumenhagen:2013fgp}. Moreover, they have played a central role in modified gravity and cosmology, giving rise to the self-accelerating solutions \cite{Dvali:2000hr, Deffayet:2000uy, Deffayet:2001pu, Deffayet:2001uk, Deffayet:2002sp}, and might explain why dark energy is so small through the \textit{dark dimension} \cite{Lust:2019zwm,Montero:2022prj, Anchordoqui:2024ajk}.

When extending theories to higher dimensions, it is important to determine which of their properties change. This is especially interesting in the context of degrees of freedom -- the building blocks of such theories -- which are essential to know how stable the theories are, and infer their observational signatures. However, while one can naturally expect that writing Einstein's gravity, a theory whose action contains only the Ricci scalar, in four or higher dimensions will preserve its properties in a straightforward way, the same is not clear if one extends the symmetry of a theory from four to d dimensions. In this proceedings, based on the work in \cite{Hell:2025wha}, we will discuss how such extensions can change fundamental properties of theories by focusing on two examples: pure, scale-invariant and conformal theories of gravity. 

In four dimensions, theories with quadratic powers of curvature play a central role among possible UV completions of gravity due to their intriguing properties  \cite{Adler:1982ri, tHooft:2011aa}. Unlike Einstein's General Relativity, they are argued to be renormalizable and asymptotically free \cite{Stelle:1976gc, Stelle:1977ry, Julve:1978xn, Fradkin:1981iu}. Higher order gravities also provide an upper bound for the effective action, since the species scale \cite{Dvali:2007hz,Dvali:2007wp,Dvali:2009ks,Dvali:2010vm,Dvali:2012uq} is controlled by the moduli dependent coefficients of higher curvature operators \cite{vandeHeisteeg:2022btw,Cribiori:2022nke,Calderon-Infante:2025ldq}. When considering  quadratic theories of gravity, among all possible combinations of terms that one can write, two symmetries play an important role. First, if one neglects the term linear in the Ricci scalar, the general action of quadratic gravity is scale invariant, meaning that it won't change upon constant rescaling of the metric. The second is conformal invariance, which can be obtained by taking a square of the Weyl-tensor, and yields a theory that is invariant under conformal transformations, thus  depending only on the angles, and not on the distances \cite{Weyl:1918ib, Weyl:1919fi}. 

In the most general case, scale-invariant gravity in four dimensions contains higher-order derivatives, leading to the appearance of the Ostrogradsky ghosts -- modes that classically lead to instability, and give rise to unitarity violation when the theory is quantized \cite{Ostrogradsky:1850fid}. In flat spacetime, these appear in both the scalar and tensor sectors yielding two scalars among which one is healthy, and one pathological, as well as two types of gravitational waves, with one being pathological. The vector modes, in contrast, are well-behaved. The situation can get somewhat better if one restricts to conformal gravity -- in this theory, the scalar modes no longer propagate. However, the tensor modes still appear with higher-order time derivatives. While the possibility of removing the pathological ghosts through boundary conditions exists \cite{Maldacena:2011mk, Anastasiou:2016jix, Anastasiou:2020mik, Hell:2023rbf}, they otherwise remain. Therefore, among all possible terms, one can only write a square of the Ricci scalar to obtain a scale-invariant, ghost-free theory of quadratic gravity. Such \textit{"pure"} theory is also interesting on its own -- as shown in \cite{Hell:2023mph} in flat spacetime it describes no propagating modes, while if the spacetime is curved, it gives rise to the (healthy) gravitational waves and a scalar field \cite{Alvarez-Gaume:2015rwa}\footnote{See also \cite{Golovnev:2024sod, Karananas:2024hoh, Barker:2025gon} for a further discussion on the degrees of freedom and the Hamiltonian formulation of the theory.}.

Higher-derivative theories of gravity are expected to naturally emerge from extra dimensional theories \cite{Fradkin:1985am, Alvarez-Gaume:2015rwa}. While at the same time keeping in mind that some of these models may have an underlying symmetry such as scale or conformal invariance, it is natural to explore if the resulting theories would keep the same properties as their four-dimensional counterparts. Following the detailed analysis of \cite{Hell:2025wha}, in these proceedings, we will discuss this question on two particular examples of higher-derivative gravities -- the pure scale-invariant gravity, and a simplest case of conformal gravity --  and demonstrate how one can deal examples of such \textit{gravitational monsters} in an elegant way. 

In particular, in the Section \ref{sec::action}, we will present the theories in d dimensions, and have a first look at their spectrum in flat spacetime. In Section \ref{sec::frames}, we will review the notion of gravitational frames, and how they can help us to analyze the spectrum of theories, applying it to both of our examples. In Section \ref{sec::alt} we will then present the summary of the results for conformal gravity in five dimensions, and finally conclude in the Section \ref{sec::con}.

\section{Conformal and scale-invariant actions in d dimensions}\label{sec::action}
Let us start our analysis by specifying the actions that are invariant under conformal and scale transformations of the metric, and having a look at their behavior in flat spacetime.  
In d dimensions, upon rescaling the metric by 
\begin{equation}
    \Tilde{g}_{\mu\nu}= F g_{\mu\nu}, 
\end{equation}
with $F$ a constant, one finds that the curvature terms transform as: 
\begin{equation}
    \Tilde{R}=\frac{1}{F}R,\qquad    \Tilde{R}_{\mu\nu}=R_{\mu\nu},\qquad \text{and},\qquad \Tilde{R}_{\mu\nu\rho\sigma}=FR_{\mu\nu\rho\sigma},
\end{equation}
where tilde denotes the curvature with respect to the rescaled metric $\Tilde{g}_{\mu\nu}$. This means the contracted square of the curvature tensors will yield a factor of $1/F^2$ in total (as in the case of four dimensions), or, more generally $1/F^d$ depending on the power for the fully contracted expression, while from the metric determinant we find a factor of $F^{d/2}$. Therefore, scale-invariant gravity can be schematically written as 
\begin{equation}
    S=\int d^dx\sqrt{-g}\sum\left(\text{curvature}\right)^\frac{d}{2},
\end{equation}
where $\text{curvature}^d$ denotes the Ricci scalar, and contractions of the Ricci tensors and the Riemann tensors. The sum represents summation of all possible combinations of these terms, such that the overall power of curvature is $d/2$. In the above, we have assumed that the action is parity even. If one instead also considers the possibility to include contractions with the Levi-Civita tensor and the curvature tensors, then one has to also take its transformation when forming the scale invariant action. 

Following \cite{Hell:2025wha}, we define \textit{pure, scale-invariant gravity}, as:
\begin{equation}\label{action_scale_inv}
    S=\beta_d\int d^dx \sqrt{-g} R^{\frac{d}{2}},
\end{equation}
such that it contains only the Ricci scalar. In the above expression, $\beta_d$ is the coupling constant.  

In contrast to scale-invariance, conformal invariance requires that the theory is invariant under the conformal transformations, for which the function now becomes dependent on the coordinates: 
\begin{equation}\label{conftrans}
    \Tilde{g}_{\mu\nu}= F\left(x^{\sigma}\right) g_{\mu\nu}.
\end{equation}
As a result, requirement that the action is invariant under conformal transformations yields a more complicated transformation law for the Riemann tensor, as well as the Ricci tensor and the Ricci scalar. However, the Weyl tensor, which is in d dimensions defined as
\begin{equation}
    \begin{split}
        W_{\mu\nu\rho\sigma}&= R_{\mu\nu\rho\sigma}-\frac{1}{d-2}\left(g_{\mu\rho}R_{\nu\sigma}-g_{\mu\sigma}R_{\nu\rho}-g_{\nu\rho}R_{\mu\sigma}+g_{\nu\sigma}R_{\mu\rho}\right)\\&+\frac{1}{(d-1)(
    d-2)}R\left(g_{\mu\rho}g_{\nu\sigma}-g_{\mu\sigma}g_{\nu\rho}\right),
    \end{split}
\end{equation}
takes a particularly simple transformation under (\ref{conftrans}): 
\begin{equation}
    \Tilde{W}_{\mu\nu\rho\sigma}=FW_{\mu\nu\rho\sigma}.
\end{equation}
Therefore, we can easily form conformally-invariant actions by contracting different powers of the Weyl-tensor, and schematically present the most general (parity-even) one as:
\begin{equation}
    S=\int d^dx\sqrt{-g}\sum\left(\text{Weyl-tensor}\right)^{\frac{d}{2}}
\end{equation}
where the expression to the power of $d$ stands for all allowed contractions of the Weyl tensor such that the action is invariant under conformal transformations, with sum over all such possibilities. For example, in 8 dimensions, we can include:
\begin{equation}
     S=\int d^8x\sqrt{-g}\left[c_1\left(W_{\mu\nu\rho\sigma}W^{\mu\nu\rho\sigma}\right)^2+c_2W_{\mu\nu\rho\sigma}W^{\rho\sigma\alpha\beta}W_{\alpha\beta\gamma\delta}W^{\gamma\delta\mu\nu}\right],
\end{equation}
where $c_1$ and $c_2$ denote the coupling constants \footnote{See also for the other conformal invariants in 8 dimensions \cite{Boulanger:2004zf, Paci:2024xxz}.}. 

In these proceedings, we will focus on the simplest conformally-invariant action, introduced in \cite{Hell:2025wha}: 
\begin{equation}\label{action_conf_inv}
    S=\alpha_{\text{CG}}\int d^dx\sqrt{-g}\sum\left(W_{\mu\nu\rho\sigma}W^{\mu\nu\rho\sigma}\right)^{\frac{d}{4}}
\end{equation}
which we will refer to as \textit{conformal gravity in d dimensions}. In the above, $\alpha_{\text{CG}}$ is the coupling constant. Clearly, in $d=4$, the above action reduces to the well-known conformal gravity in four dimensions.

Let us now consider the content of the pure, scale-invariant, and conformal actions in the simplest possible background  -- the Minkowski spacetime. One might naively think that extending a description from four to d dimensions is straightforward, but we can notice an obstacle right away: the actions (\ref{action_scale_inv}) and (\ref{action_conf_inv}) are non-polynomial in this case. Luckily, the pure scale-invariant case is still not too complicated. By perturbing the metric around the flat spacetime
\begin{equation}
    g_{\mu\nu}=\eta_{\mu\nu}+h_{\mu\nu}
\end{equation}
and decomposing the metric perturbations according to the spatial rotations, 
\begin{equation}\label{ddimdec}
    \begin{split}
        h_{00}&=2\phi\\
        h_{0i}&=S_i+B_{,i}\\
        h_{ij}&=2\psi\delta_{ij}+E_{,ij}+F_{i,j}+F_{j,i}+h^T_{ij}, 
    \end{split}
\end{equation}
where $i=1,...,d-1$, and vector and tensor modes satisfy:
\begin{equation}
    S_{i,i}=0\qquad F_{i,i}=0\qquad h_{ij,i}=0\qquad h_{ii}=0. 
\end{equation}
we find that at the leading order, in the Newtonian gauge generalization to d dimensions\footnote{In four dimensions, the scalars $\phi$ and $\psi$ are known as the Bardeen potentials \cite{Bardeen:1980kt, Mukhanov:1990me}.} with $B=0$ and $E=0$, the action is only a function of the two (gauge-invariant) scalar modes:
\begin{equation}
    S=\beta_d\int d^dx 2^{d/2}(\Delta\phi+\Ddot{\psi}-2\Delta\psi)^{d/2}
\end{equation}
However, $\phi$ does not propagate, but rather leads to the constraint
\begin{equation}
    \Delta\left[\left(\Delta\phi+\Ddot{\psi}-2\Delta\psi \right)^{\frac{d-2}{d}}\right]=0,
\end{equation}
which when substituted back into the action sets 
\begin{equation}
    S=0.
\end{equation}
Thus, similarly to the pure scale-invariant gravity in four dimensions \cite{Hell:2023mph}, the d-dimensional generalization propagates no modes in flat spacetime, at least in the leading order in the perturbation theory, indicating that the Minkowski background is rigid. 

Unfortunately, the simplest action for conformal gravity (\ref{action_conf_inv}) is not so easy for the same background. In contrast to the pure, scale-invariant case, scalar, vector and tensor perturbations now mix, therefore giving rise to much more complicated constraints. For the conformally invariant background 
\begin{equation}
    ds^2=a^2(\eta)\eta_{\mu\nu}dx^{\mu}dx^{\nu}
\end{equation}
the leading-order Lagrangian density corresponding to the action (\ref{action_conf_inv}) takes the following form: 
\begin{equation}
\mathcal{L}=\alpha_{\text{CG}}\left[P_S+P_V+P_T+P_{SV}+P_{ST}+P_{VT}\right]^{\frac{d}{4}},
\end{equation}
where,
\begin{equation*}
    \begin{split}
      P_S&=\frac{4(d-3)}{d-2}\chi_{,ij}\chi_{,ij}-\frac{4(d-3)}{(d-1)(d-2)}\Delta\chi\Delta\chi\\
      P_V&=\frac{2(d-3)}{d-2}(\dot{S}_{i,j}\dot{S}_{i,j}+\dot{S}_{i,j}\dot{S}_{j,i})+2S_{i,jk}S_{k,ij}-2S_{k,ij}S_{k,ij}+\frac{2}{d-2}\Delta S_i\Delta S_i\\
      P_T&=h_{ij,\mu\nu}^Th_{ij}^{T,\mu\nu}-\frac{1}{d-2}\Box h_{ij}^T\Box h_{ij}^T+2(\dot{h}_{ij,k}^T\dot{h}_{jk,i}-h_{ij,kl}^Th_{jl,ik}^T)+h_{ij,kl}^Th_{kl,ij}^T\\
      P_{SV}&=-\frac{8(d-3)}{d-2}\dot{S}_{j,i}\chi_{,ij}\\
      P_{ST}&=\frac{4(d-3)}{d-2}\Ddot{h}_{ij}^T\chi_{,ij}+\frac{4}{d-2}\Delta h_{ij}^T\chi_{,ij}\\
      P_{VT}&=-2(\dot{S}_{i,j}\Ddot{h}_{ij}^T-S_{i,kj}\dot{h}_{ik,j}+\dot{S}_{j,i}\Ddot{h}_{ij}^T+S_{i,kl}\dot{h}_{kl,i}^T-2S_{k,ij}\dot{h}_{ij,k}^T)-\frac{4}{d-2}\dot{S}_{j,i}\Box h_{ij}^T,
    \end{split}
\end{equation*}
and where the dot stands for the derivative with respect to the conformal time. 
To obtain the above expression, we have decomposed the metric according to the spatial rotations, implemented the Newtonian gauge, and have defined $\chi=\phi+\psi$. Since $\chi$ is a non-propagating field, it indicates that the theory might have the same number of degrees of freedom, at least for the conformally invariant background, matching the four-dimensional case. However, due to the mixing between scalars, vectors, and tensors, clearly, it is non-trivial to express the action only in terms of the vector and tensor modes. Because of the conformal invariance, the theory will be non-polynomial for any conformally-invariant background. Therefore, to obtain a better understanding of its content, in the following sections, we will consider other types of background, and for such, rewrite the action in a clearer way.

\section{Frames for pure, scale-invariant and conformal gravities}\label{sec::frames}

The notion of frames in gravity is particularly convenient when considering theories beyond Einstein's relativity. It is a way to bring the theory that is initially presented in a complicated form, to an expression that is easy to analyse. One of the well-known examples of such theories is the $f(R)$ gravity. In its original form, written in \textit{the Jordan frame}, the action can be analyzed, but clearly in a more difficult way when compared to just General Relativity. However, by introducing a scalar field in a special way and performing a conformal transformation, one writes the same theory such that the resulting action becomes linear in the Ricci scalar, containing in addition a scalar field with a potential  \cite{Teyssandier:1983zz, Whitt:1984pd, Barrow:1988xh, Wands:1993uu}. In this form, the theory is said to be written in \textit{the Einstein frame}. 

Following this notion, let us also write the actions for scale-invariant and conformal gravities in a way that allows us to infer their properties more easily. 

The trick for the pure scale-invariant gravity is similar to its four-dimensional case \cite{Alvarez-Gaume:2015rwa}. Starting with the pure scale-invariant action in d dimensions (\ref{action_scale_inv}), we can introduce a new scalar field: 
\begin{equation}
    S=\frac{d(d-2)}{4}\int d^dx\sqrt{-g}\left[\frac{2}{d-2}R\phi^{\frac{d-2}{2}}-\frac{2}{d}\beta_d^{\frac{-2}{d-2}}\phi^{\frac{d}{2}}\right], 
\end{equation}
which is related to the Ricci scalar through a constraint: 
\begin{equation}
    \phi=\beta_d^{\frac{2}{d-2}}R. 
\end{equation}
If one substitutes this constraint back to the action, one can recover (\ref{action_scale_inv}). However, since our goal is to write (\ref{action_scale_inv}) in the Einstein frame, we will instead perform conformal transformation, defining a new metric: 
\begin{equation}
    \Tilde{g}_{\mu\nu}=\frac{\phi}{M_P^2}g_{\mu\nu}. 
\end{equation}
By expressing the action in terms of the new metric, and by redefining 
\begin{equation}
    \Phi=\frac{\sqrt{d(d-1)(d-2)}}{2M_P}\ln\phi, 
\end{equation}
the action (\ref{action_scale_inv}) becomes: 
\begin{equation}
    S=\int d^dx\sqrt{-\Tilde{g}}\left[\frac{d M_P^2}{2}\Tilde{R}-\frac{1}{2}\Tilde{\nabla}_{\mu}\Phi\Tilde{\nabla}^{\mu}\Phi-dM_P^{d-2}\beta_d^{\frac{-2}{d-2}}\Lambda\right], 
\end{equation}
where 
\begin{equation}
    \Lambda=\frac{d-2}{2}M_P^2\beta_d^{\frac{-2}{d-2}}, 
\end{equation}
and $\Tilde{R}$ and $\Tilde{\nabla}$ are the Ricci scalar and the covariant derivative with respect to the new metric respectively. Therefore, we can see that for a general curved background, the pure, scale-invariant gravity will describe a graviton and a scalar mode. However, this description is only valid for backgrounds where $R\neq 0$. Otherwise, the transformation between the frames becomes singular, and one should use the original frame to analyze theory. This change in the degrees of freedom between the two frames can be also found in more general frameworks, such as non-minimal couplings between a scalar field and curvature \cite{Hell:2025lbl}.  

Up to the singular cases, we can notice that the change in frame greatly simplifies the description of pure scale-invariant gravity. Following \cite{Hell:2025wha}, we can use the similar trick to also simplify the action for the conformal gravity, and define \textit{the alternative frame}. In order to do this, we introduce again a scalar field:
\begin{equation}\label{cg_act_2}
    S=\alpha_{\text{CG}}\int d^dx\sqrt{-g}\left[\frac{d}{4}W_{\mu\nu\rho\sigma}W^{\mu\nu\rho\sigma}\phi^{\frac{d-4}{4}}-\frac{d-4}{4}\phi^{\frac{d}{4}}\right],
\end{equation}
which satisfies the constraint: 
\begin{equation}
    \phi=W_{\mu\nu\rho\sigma}W^{\mu\nu\rho\sigma}. 
\end{equation}
By substituting this constraint back to the action, one can recover the original theory (\ref{action_conf_inv}). However, if one instead performs conformal transformation: 
\begin{equation}
    \Tilde{g}_{\mu\nu}=\frac{\sqrt{\phi}}{M_P^2}g_{\mu\nu}, 
\end{equation}
the action (\ref{cg_act_2}) becomes: 
\begin{equation}\label{d=5act}
    S=\int d^dx \sqrt{-\Tilde{g}}\left[\frac{\alpha_{\text{CG}}dM_P^{d-4}d}{4}\Tilde{W}_{\mu\nu\rho\sigma}\Tilde{W}^{\mu\nu\rho\sigma}-M_P^{d-2}\Lambda\right], 
\end{equation}
where 
\begin{equation}
    \Lambda=M_P^2\alpha_{\text{CG}}\frac{d-4}{4}
\end{equation}
is the cosmological constant, and $\Tilde{W}_{\mu\nu\rho\sigma}$ is the Weyl tensor with respect  to the new metric. One should note however that the above action is valid only for backgrounds that are not conformally invariant, since otherwise the Weyl tensor vanishes. However, it still provides a more elegant way to infer the nature and number of the degrees of freedom. It is important to note that conformal invariance is not lost in the alternative frame. Rather, the new metric itself is already conformally invariant, which makes it particularly interesting to impose boundary conditions to eliminate ghost modes, which were in four dimensions explicitly breaking the conformal symmetry \cite{Maldacena:2011mk, Anastasiou:2016jix, Anastasiou:2020mik, Hell:2023rbf}.

\section{The ghosts in five dimensions}\label{sec::alt}
Let us now see how the structure of conformal gravity changes in five dimensions, when compared to its four dimensional analogue. The alternative frame is particularly useful for this, but it requires a background that is not conformally invariant. We will therefore consider the anisotropic spacetime, whose background is given by: 
\begin{equation}\label{anis5}
    ds^2=-dt^2+a^2(t)\delta_{ij}dx^idx^j+b^2(t)dl^2.
\end{equation}
Here, $a$ and $b$ are the scale factors that only depend on time, and the index i has values 1, 2 and 3. As shown in \cite{Hell:2025wha}, the action (\ref{d=5act}) with $d=5$ has both analytical and numerical solutions for the above anisotropic background, admitting accelerating stage, as well as transitions from decelerated, to super-accelerated, and de Sitter stage phase of expansion. The particular choice of the anisotropy also makes it convenient to analyze the propagating modes. In particular, we can apply the scalar-vector-tensor decomposition \cite{vandeBruck:2000ju} for the metric perturbations $\delta g_{\mu\nu}$ around the anisotropic background $ g_{\mu\nu}= g_{\mu\nu}^{(0)}+\delta g_{\mu\nu}$:
\begin{equation}
    \begin{split}
        &\delta g_{00}=2\phi\\
        &\delta g_{oi}=S_i+B_{,i},\qquad S_{i,i}=0\\
         &\delta g_{0l}=\omega_{,l}\\
        &\delta g_{ij}=a^2(2\psi \delta_{ij}+2E_{,ij}+F_{i,j}+F_{j,i}+h_{ij}^T), \qquad F_{i,i}=0,\qquad h^T_{ii}=0,\qquad h^T_{ij,i}=0\\
         &\delta g_{ll}=b^2\sigma\\
        &\delta g_{li}=v_{i,l}+\mu_{,li}\qquad v_{i,i}=0,
    \end{split}
\end{equation}
with $i$ accounting for the $x,y,$ and $z$ coordinates. The above can be further simplified by choosing a gauge \begin{equation}
    E=0\qquad B=0\qquad\mu=0\qquad F_i=0, 
\end{equation}
in which remaining components match the gauge-invariant variables. Moreover, by expanding the action to terms quadratic in the perturbations, one can notice that similarly to the standard cosmological perturbation theory, scalar, vector and tensor modes decouple from each other. 

To find the resulting modes, one can follow the procedure described in \cite{Hell:2025wha}, or to express the action in terms of the propagating modes, by following \cite{Hell:2026blj}, reduce the higher-order derivatives to second order, find the constraints, solve, and substitute back to the action. Both procedures give rise to the same results, which we will summarize in the following, and compare to the four dimensional case:

\textbf{The scalar sector} contains three propagating modes, one of which is a ghost in five dimensions. The four-dimensional case in contrast contains no scalar modes. 

\textbf{The vector sector} also contains three types of propagating vector modes, hence giving rise to six additional degrees of freedom, due to them being transverse. One among these three vectors is also a ghost, corresponding to two ghost degrees of freedom. This is in contrast with the four-dimensional case, which has two healthy degrees of freedom, corresponding to one propagating vector. 

\textbf{The tensor sector} remains the same as its four-dimensional analogue, having a ghost tensor, and a healthy one, thus giving rise to four additional degrees of freedom. 

Therefore, overall, we can see that conformal gravity has a larger field content in five dimensions when compared to the four-dimensional case. This suggests that the general scale-invariant gravity could also increase its spectrum -- while in four dimensions it has two scalar modes, and one vector mode, in higher dimensions it could propagate more modes, and have thus more ghost degrees of freedom, unless one finds a very particular choice that could potentially cancel them. In other words, it appears that only pure scale-invariant case keeps its properties in four or higher dimensions.

\section{Conclusion}\label{sec::con}
It is well-known that gravity changes its properties in two, three or four dimensions. For example, two or three dimensions do not admit gravitational waves. However, naively, one might suspect that generalization to dimensions that are larger than four is, in contrast, trivial. In these proceedings, we have explored this question in two types of gravitational theories -- the scale invariant gravity, with a special focus on the pure case, and gravity with conformal invariance, comparing them to their four-dimensional analogues. 

The pure, scale-invariant gravity, which is a function only of the Ricci scalar, takes a special place among all possible choices of higher-curvature gravity. In both four and higher dimensions it describes no propagating modes for flat spacetime, and otherwise describes a graviton and a scalar field. 
However, the similarity between higher-curvature gravity in four or higher dimensions ends with this special example. By studying conformal gravity in $d> 4$, we found that the spectrum entirely changes. Moreover, the theory is particularly difficult to analyze, even at the leading order in the perturbations. The main reason for this is the non-polynomial nature of the theory, which mixes scalar, vector and tensor modes for conformally-flat spacetimes. 

To counter this issue, and obtain a deeper insight into the properties of the theory, we have introduced an alternative frame, in which the theory can be written in terms of the square of the Weyl tensor, together with a cosmological constant. While this form is not applicable for conformally invariant background solutions, it is at the same time especially nice because the corresponding metric is conformally invariant by construction. 

However, by further studying this theory in five-dimensional anisotropic spacetime, we have found that the theory describes a significantly larger and more problematic spectrum, than its four-dimensional version. In particular, the ghost modes, which were previously contained only in the tensor sector, now also appear for the vector and scalar modes. In addition, there are also healthy modes that are not present in four-dimensional conformal gravity -- the theory contains two types of healthy vectors and scalars, whereas the four-dimensional case has only one type of vector and no scalar modes. 

Overall, up to special examples, d-dimensional extensions of higher-derivative gravity radically change the propagating modes. The increase of Ostrogradsky ghosts in conformal gravity suggests that these pathologies will remain even in the general, scale-invariant case.

\section*{Acknowledgments}

A. H. would like to thank CORFU2025 organizers for the invitation to present this work. The work of A. H. was supported by the World Premier International Research Center Initiative (WPI), MEXT, Japan, and in part by JSPS KAKENHI No. JP24K00624. The work of D.L. is supported by the German-Israel-Project (DIP) on Holography and the Swampland.


\begin{thebibliography}{99}

\bibitem{Kaluza:1921tu}
T.~Kaluza,
Sitzungsber. Preuss. Akad. Wiss. Berlin (Math. Phys. ) \textbf{1921}, 966-972 (1921)
doi:10.1142/S0218271818700017
[arXiv:1803.08616 [physics.hist-ph]].

\bibitem{Klein:1926tv}
O.~Klein,
Z. Phys. \textbf{37}, 895-906 (1926)
doi:10.1007/BF01397481

\bibitem{Arkani-Hamed:1998sfv}
N.~Arkani-Hamed, S.~Dimopoulos and G.~R.~Dvali,
Phys. Rev. D \textbf{59}, 086004 (1999)
doi:10.1103/PhysRevD.59.086004
[arXiv:hep-ph/9807344 [hep-ph]].

\bibitem{Arkani-Hamed:1998jmv}
N.~Arkani-Hamed, S.~Dimopoulos and G.~R.~Dvali,
Phys. Lett. B \textbf{429}, 263-272 (1998)
doi:10.1016/S0370-2693(98)00466-3
[arXiv:hep-ph/9803315 [hep-ph]].

\bibitem{Antoniadis:1998ig}
I.~Antoniadis, N.~Arkani-Hamed, S.~Dimopoulos and G.~R.~Dvali,
Phys. Lett. B \textbf{436}, 257-263 (1998)
doi:10.1016/S0370-2693(98)00860-0
[arXiv:hep-ph/9804398 [hep-ph]].

\bibitem{Randall:1999ee}
L.~Randall and R.~Sundrum,
Phys. Rev. Lett. \textbf{83}, 3370-3373 (1999)
doi:10.1103/PhysRevLett.83.3370
[arXiv:hep-ph/9905221 [hep-ph]].

\bibitem{Randall:1999vf}
L.~Randall and R.~Sundrum,
Phys. Rev. Lett. \textbf{83}, 4690-4693 (1999)
doi:10.1103/PhysRevLett.83.4690
[arXiv:hep-th/9906064 [hep-th]].

\bibitem{Green:1987sp}
M.~B.~Green, J.~H.~Schwarz and E.~Witten,
1988,
ISBN 978-0-521-35752-4

\bibitem{Blumenhagen:2013fgp}
R.~Blumenhagen, D.~L{\"u}st and S.~Theisen,
Springer, 2013,
ISBN 978-3-642-29496-9
doi:10.1007/978-3-642-29497-6

\bibitem{Dvali:2000hr}
G.~R.~Dvali, G.~Gabadadze and M.~Porrati,
Phys. Lett. B \textbf{485}, 208-214 (2000)
doi:10.1016/S0370-2693(00)00669-9
[arXiv:hep-th/0005016 [hep-th]].

\bibitem{Deffayet:2000uy}
C.~Deffayet,
Phys. Lett. B \textbf{502}, 199-208 (2001)
doi:10.1016/S0370-2693(01)00160-5
[arXiv:hep-th/0010186 [hep-th]].

\bibitem{Deffayet:2001pu}
C.~Deffayet, G.~R.~Dvali and G.~Gabadadze,
Phys. Rev. D \textbf{65}, 044023 (2002)
doi:10.1103/PhysRevD.65.044023
[arXiv:astro-ph/0105068 [astro-ph]].

\bibitem{Deffayet:2001uk}
C.~Deffayet, G.~R.~Dvali, G.~Gabadadze and A.~I.~Vainshtein,
Phys. Rev. D \textbf{65}, 044026 (2002)
doi:10.1103/PhysRevD.65.044026
[arXiv:hep-th/0106001 [hep-th]].

\bibitem{Deffayet:2002sp}
C.~Deffayet, S.~J.~Landau, J.~Raux, M.~Zaldarriaga and P.~Astier,
Phys. Rev. D \textbf{66}, 024019 (2002)
doi:10.1103/PhysRevD.66.024019
[arXiv:astro-ph/0201164 [astro-ph]].

\bibitem{Lust:2019zwm}
D.~L{\"u}st, E.~Palti and C.~Vafa,
Phys. Lett. B \textbf{797}, 134867 (2019)
doi:10.1016/j.physletb.2019.134867
[arXiv:1906.05225 [hep-th]].

\bibitem{Montero:2022prj}
M.~Montero, C.~Vafa and I.~Valenzuela,
JHEP \textbf{02}, 022 (2023)
doi:10.1007/JHEP02(2023)022
[arXiv:2205.12293 [hep-th]].

\bibitem{Anchordoqui:2024ajk}
L.~A.~Anchordoqui, I.~Antoniadis and D.~L{\"u}st,
PoS \textbf{CORFU2023}, 215 (2024)
doi:10.22323/1.463.0215
[arXiv:2405.04427 [hep-th]].

\bibitem{Hell:2025wha}
A.~Hell and D.~L{\"u}st,
JHEP \textbf{09} (2025), 202
doi:10.1007/JHEP09(2025)202
[arXiv:2506.18775 [hep-th]].

\bibitem{Adler:1982ri}
S.~L.~Adler,
Rev. Mod. Phys. \textbf{54}, 729 (1982)
[erratum: Rev. Mod. Phys. \textbf{55}, 837 (1983)]
doi:10.1103/RevModPhys.54.729

\bibitem{tHooft:2011aa}
G.~'t Hooft,
Found. Phys. \textbf{41}, 1829-1856 (2011)
doi:10.1007/s10701-011-9586-8
[arXiv:1104.4543 [gr-qc]].

\bibitem{Stelle:1976gc}
K.~S.~Stelle,
Phys. Rev. D \textbf{16}, 953-969 (1977)
doi:10.1103/PhysRevD.16.953

\bibitem{Stelle:1977ry}
K.~S.~Stelle,
Gen. Rel. Grav. \textbf{9}, 353-371 (1978)
doi:10.1007/BF00760427

\bibitem{Julve:1978xn}
J.~Julve and M.~Tonin,
Nuovo Cim. B \textbf{46}, 137-152 (1978)
doi:10.1007/BF02748637

\bibitem{Fradkin:1981iu}
E.~S.~Fradkin and A.~A.~Tseytlin,
Nucl. Phys. B \textbf{201}, 469-491 (1982)
doi:10.1016/0550-3213(82)90444-8

\bibitem{Dvali:2007hz}
G.~Dvali,
Fortsch. Phys. \textbf{58}, 528-536 (2010)
doi:10.1002/prop.201000009
[arXiv:0706.2050 [hep-th]].

\bibitem{Dvali:2007wp}
G.~Dvali and M.~Redi,
Phys. Rev. D \textbf{77}, 045027 (2008)
doi:10.1103/PhysRevD.77.045027
[arXiv:0710.4344 [hep-th]].

\bibitem{Dvali:2009ks}
G.~Dvali and D.~L{\"u}st,
Fortsch. Phys. \textbf{58}, 505-527 (2010)
doi:10.1002/prop.201000008
[arXiv:0912.3167 [hep-th]].

\bibitem{Dvali:2010vm}
G.~Dvali and C.~Gomez,
[arXiv:1004.3744 [hep-th]].

\bibitem{Dvali:2012uq}
G.~Dvali, C.~Gomez and D.~L{\"u}st,
Fortsch. Phys. \textbf{61}, 768-778 (2013)
doi:10.1002/prop.201300002
[arXiv:1206.2365 [hep-th]].

\bibitem{vandeHeisteeg:2022btw}
D.~van de Heisteeg, C.~Vafa, M.~Wiesner and D.~H.~Wu,
Beijing J. Pure Appl. Math. \textbf{1}, no.1, 1-41 (2024)
doi:10.4310/bpam.2024.v1.n1.a1
[arXiv:2212.06841 [hep-th]].

\bibitem{Cribiori:2022nke}
N.~Cribiori, D.~L{\"u}st and G.~Staudt,
Phys. Lett. B \textbf{844}, 138113 (2023)
doi:10.1016/j.physletb.2023.138113
[arXiv:2212.10286 [hep-th]].

\bibitem{Calderon-Infante:2025ldq}
J.~Calder{\'o}n-Infante, A.~Castellano and A.~Herr{\'a}ez,
SciPost Phys. \textbf{19}, no.4, 096 (2025)
doi:10.21468/SciPostPhys.19.4.096
[arXiv:2501.14880 [hep-th]].

\bibitem{Weyl:1918ib}
H.~Weyl,
Sitzungsber. Preuss. Akad. Wiss. Berlin (Math. Phys. ) \textbf{1918}, 465 (1918)


\bibitem{Weyl:1919fi}
H.~Weyl,
Annalen Phys. \textbf{59}, 101-133 (1919)
doi:10.1002/andp.19193641002

\bibitem{Ostrogradsky:1850fid}
M.~Ostrogradsky,
Mem. Acad. St. Petersbourg \textbf{6}, no.4, 385-517 (1850)

\bibitem{Maldacena:2011mk}
J.~Maldacena,
[arXiv:1105.5632 [hep-th]].

\bibitem{Anastasiou:2016jix}
G.~Anastasiou and R.~Olea,
Phys. Rev. D \textbf{94}, no.8, 086008 (2016)
doi:10.1103/PhysRevD.94.086008
[arXiv:1608.07826 [hep-th]].

\bibitem{Anastasiou:2020mik}
G.~Anastasiou, I.~J.~Araya and R.~Olea,
JHEP \textbf{01}, 134 (2021)
doi:10.1007/JHEP01(2021)134
[arXiv:2010.15146 [hep-th]].

\bibitem{Hell:2023rbf}
A.~Hell, D.~L{\"u}st and G.~Zoupanos,
JHEP \textbf{08}, 168 (2023)
doi:10.1007/JHEP08(2023)168
[arXiv:2306.13714 [hep-th]].

\bibitem{Hell:2023mph}
A.~Hell, D.~L{\"u}st and G.~Zoupanos,
JHEP \textbf{02}, 039 (2024)
doi:10.1007/JHEP02(2024)039
[arXiv:2311.08216 [hep-th]].



\bibitem{Alvarez-Gaume:2015rwa}
L.~Alvarez-Gaume, A.~Kehagias, C.~Kounnas, D.~L{\"u}st and A.~Riotto,
Fortsch. Phys. \textbf{64}, no.2-3, 176-189 (2016)
doi:10.1002/prop.201500100
[arXiv:1505.07657 [hep-th]].
 
\bibitem{Golovnev:2024sod}
A.~Golovnev,
Ukr. J. Phys. \textbf{69}, no.7, 456 (2024)
doi:10.15407/ujpe69.7.456
[arXiv:2405.14184 [gr-qc]].

\bibitem{Karananas:2024hoh}
G.~K.~Karananas,
Phys. Rev. D \textbf{111}, no.4, 044068 (2025)
doi:10.1103/PhysRevD.111.044068
[arXiv:2407.09598 [hep-th]].

\bibitem{Barker:2025gon}
W.~Barker and D.~Glavan,
[arXiv:2510.08201 [gr-qc]].

\bibitem{Fradkin:1985am}
E.~S.~Fradkin and A.~A.~Tseytlin,
Phys. Rept. \textbf{119}, 233-362 (1985)
doi:10.1016/0370-1573(85)90138-3

\bibitem{Bardeen:1980kt}
J.~M.~Bardeen,
Phys. Rev. D \textbf{22}, 1882-1905 (1980)
doi:10.1103/PhysRevD.22.1882

\bibitem{Mukhanov:1990me}
V.~F.~Mukhanov, H.~A.~Feldman and R.~H.~Brandenberger,
Phys. Rept. \textbf{215}, 203-333 (1992)
doi:10.1016/0370-1573(92)90044-Z

\bibitem{Teyssandier:1983zz}
P.~Teyssandier and P.~Tourrenc,
J. Math. Phys. \textbf{24}, 2793 (1983)
doi:10.1063/1.525659

\bibitem{Whitt:1984pd}
B.~Whitt,
Phys. Lett. B \textbf{145}, 176-178 (1984)
doi:10.1016/0370-2693(84)90332-0

\bibitem{Barrow:1988xh}
J.~D.~Barrow and S.~Cotsakis,
Phys. Lett. B \textbf{214}, 515-518 (1988)
doi:10.1016/0370-2693(88)90110-4

\bibitem{Wands:1993uu}
D.~Wands,
Class. Quant. Grav. \textbf{11}, 269-280 (1994)
doi:10.1088/0264-9381/11/1/025
[arXiv:gr-qc/9307034 [gr-qc]].

\bibitem{Hell:2025lbl}
A.~Hell and D.~L{\"u}st,
JHEP \textbf{12}, 091 (2025)
doi:10.1007/JHEP12(2025)091
[arXiv:2509.20217 [hep-th]].

\bibitem{Boulanger:2004zf}
N.~Boulanger and J.~Erdmenger,
Class. Quant. Grav. \textbf{21} (2004) no.18, 4305-4316
[erratum: Class. Quant. Grav. \textbf{39} (2022) no.3, 039501]
doi:10.1088/0264-9381/21/18/003
[arXiv:hep-th/0405228 [hep-th]].

\bibitem{Paci:2024xxz}
G.~Paci and O.~Zanusso,
JHEP \textbf{03} (2025), 111
doi:10.1007/JHEP03(2025)111
[arXiv:2411.03842 [hep-th]].


\bibitem{vandeBruck:2000ju}
C.~van de Bruck, M.~Dorca, R.~H.~Brandenberger and A.~Lukas,
Phys. Rev. D \textbf{62}, 123515 (2000)
doi:10.1103/PhysRevD.62.123515
[arXiv:hep-th/0005032 [hep-th]].




\bibitem{Hell:2026blj}
A.~Hell, E.~G.~M.~Ferreira, D.~L{\"u}st and M.~Sasaki,
JHEP \textbf{03}, 235 (2026)
doi:10.1007/JHEP03(2026)235
[arXiv:2601.10288 [hep-th]].



\end{thebibliography}
\end{document}